# Infrared Optical Absorption in Low-spin $Fe^{2+}$-doped $SrTiO_3$

**Ryan B. Comes*,[1] Tiffany C. Kaspar,[1] Steve M. Heald,[2] Mark E. Bowden,[3] and Scott A. Chambers*[1]**

[1]Physical and Computational Sciences Directorate, Pacific Northwest National Laboratory, 902 Battelle Blvd, Richland, WA 99352, United States

[2]Advanced Photon Source, Argonne National Laboratory, 9700 S. Cass Ave, Argonne, IL 60439, United States

[3]Environmental Molecular Sciences Laboratory, Pacific Northwest National Laboratory, 902 Battelle Blvd, Richland, WA 99352, United States

E-mail: ryan.comes@pnnl.gov; sa.chambers@pnnl.gov

**Abstract.** Band gap engineering in $SrTiO_3$ and related titanate perovskites has long been explored due to the intriguing properties of the materials for photocatalysis and photovoltaic applications. A popular approach in the materials chemistry community is to substitutionally dope aliovalent transition metal ions onto the B site in the lattice to alter the valence band. However, in such a scheme there is limited control over the dopant valence, and compensating defects often form. Here we demonstrate a novel technique to controllably synthesize $Fe^{2+}$- and $Fe^{3+}$-doped $SrTiO_3$ thin films without formation of compensating defects by co-doping with $La^{3+}$ ions on the A site. We stabilize $Fe^{2+}$-doped films by doping with two La ions for every Fe dopant, and find that the Fe ions exhibit a low-spin electronic configuration, producing optical transitions in the near infrared regime and degenerate doping. The novel electronic states observed here offer a new avenue for band gap engineering in perovskites for photocatalytic and photovoltaic applications.

1. **Introduction**

SrTiO$_3$ (STO) is a commonly studied perovskite oxide of considerable interest due to its wide range of functionalities. STO and related titanates such as PbTiO$_3$ (PTO) and BaTiO$_3$ (BTO) all exhibit ferroelectricity in various forms.[1–3] This feature makes titanates intriguing candidates for ferroelectric photovoltaic applications.[3,4] However, the wide optical band gap of these materials (~3.25 eV indirect gap in STO)[5] makes them ill-suited for visible light absorption. STO also represents a good candidate for solar hydrolysis due to favorable alignment of the Ti *3d* conduction band with the half-cell potential for the H$_2$O→H$_2$ reaction.[6] Again, however, the wide band gap limits its efficiency for visible light photocatalysis. To overcome this limitation, groups have long explored ways to reduce the band gap by doping transition metal cations in place of Ti$^{4+}$. Cr$^{3+}$ is a commonly employed dopant and has been used to effectively enhance visible light photocatalysis due to the Cr *3d* electrons which lie at the top of the valence band in Cr-doped STO.[7] Through the use of a La co-dopant to stabilize Cr$^{3+}$ dopants in STO without the creation of compensating defects (such as oxygen vacancies), the direct band gap can be reduced to less than 2.5 eV in both epitaxial films and powders.[8–11] When co-doped, La$^{3+}$ substitutes for Sr$^{2+}$ and donates an electron to Cr$_{Ti}$ to stabilize Cr$^{3+}$. The stoichiometry for these materials is thus Sr$_{1-x}$La$_x$Ti$_{1-x}$Cr$_x$O$_3$ (SLTCO), since each La donor stabilizes one Cr acceptor. This co-doping scheme could in principle also be employed to stabilize Fe at the B site. By tuning the ratio of La donors to Fe acceptors, it may be possible to control the Fe valence without creating oxygen vacancies.

Recent work examining epitaxial perovskite multilayer films can provide a model for what to expect in these systems. Multilayer films comprised of LaTiO$_3$ (LTO) and LaFeO$_3$ (LFO) were recently shown to facilitate a charge transfer across the interface from Ti$^{3+}$ to Fe$^{3+}$ such that the interfacial Ti ions can adopt the more energetically favorable Ti$^{4+}$ state, donating an electron to produce Fe$^{2+}$.[12] Similar density functional theory (DFT) predictions have been made for YTiO$_3$ (YTO)-YFeO$_3$ (YFO) superlattices, with a prediction of electron transfer from Ti to Fe.[13] In both of these works, the Fe ions are predicted to take on a low-spin magnetic configuration of $(t_{2g})^6(e_g)^0$, producing no net magnetic moment on the

Fe ions. This runs counter to the naturally observed FeO wüstite phase, which has a high-spin electronic configuration of $(t_{2g})^4(e_g)^2$.[14] According to DFT models of the LTO-LFO[12] and YTO-YFO[13] super-lattices, the Fe $t_{2g}$ states in the low-spin configuration lie at energies greater than that of the O *2p* band and are very close to the Fermi level. Meanwhile, the nominally unoccupied Ti *3d* states lie at energies just a few tenths of an eV above the Fermi level. By stabilizing $Fe^{2+}$ dopants within STO in the same manner as was done in SLTCO, a band gap in the infrared regime could be achieved, complementing the visible light absorption in SLTCO to permit broad-spectrum absorption in doped titanates.

In this work, we explore the possibility of stabilizing $Fe^{2+}$ dopants in STO through co-doping with La. We demonstrate the deposition of films with formulae $Sr_{1-x}La_xTi_{1-x}Fe_xO_3$ to stabilize $Fe^{3+}$ and $Sr_{1-2x}La_{2x}Ti_{1-x}Fe_xO_3$ to stabilize $Fe^{2+}$. We will refer to these samples as $Fe^{3+}$-doped and $Fe^{2+}$-doped, respectively, though the Fe valence is not known *a priori*. Using x-ray photoelectron spectroscopy (XPS) and x-ray absorption near edge spectroscopy (XANES), we show systematic control of the valence of the Fe dopants. Furthermore, we measure the Fe-O bond length to determine the Fe dopant spin state via extended x-ray absorption fine structure (EXAFS) analysis. Valence band XPS measurements show that $Fe^{2+}$ dopants produce electronic states above the O *2p* band edge and near the Fermi level, consistent with previous DFT predictions. Spectroscopic ellipsometry measurements demonstrate that $Fe^{2+}$-doped films exhibit optical absorption in the near infrared regime with a direct bandgap for Fe $3d \rightarrow$ Ti $3d$ transitions at ~0.2 eV. Temperature-dependent resistivity measurements show that the $Fe^{2+}$ ions act as degenerate dopants near room temperature, with metallic behavior.

2. **Methods**

All films were grown on STO and $DyScO_3$ (DSO) substrates. DSO was chosen for its good lattice match (pseudo-cubic lattice parameter of 3.945 Å) to STO (a = 3.905 Å) and wide optical band gap, which enables spectroscopic ellipsometry measurements. Films were grown using a shuttered oxygen-assisted molecular beam epitaxy scheme described previously.[8,15] The samples were between 15 and 25 nm thick and were capped with 2 unit cells of STO to limit over-oxidation of the dopants during cooldown. The La to Fe ratio was calibrated using a quartz crystal oscillator such that the concentration of Fe was slightly greater than that of La (<5% greater for a ratio of 0.95 to 0.99) to avoid unintentional *n*-type doping of the films that has been observed previously in SLTCO.[8] Details of the growth process and representative reflection high-energy electron diffraction patterns and x-ray diffraction scans for the samples are shown in the supplemental information.

After growth, the samples were transferred under ultra-high vacuum conditions to an appended XPS system equipped with a monochromatic Al Kα x-ray source and Scienta R3000 analyzer for characterization. It should be noted that due to the insulating nature of the substrates, a compensating electron flood gun was used to neutralize sample charging during the measurements. Thus, the absolute binding energy cannot be determined from the measurements. However, by aligning the O 1*s* core level to 530.3 eV[16] for each sample, we can achieve reasonable estimates for the various binding energies that are consistent with published values for the various oxidation states.

Fe *K*-edge x-ray absorption spectroscopy measurements in fluorescence mode were performed at the Advanced Photon Source Sector 20-BM on samples grown on STO. These bulk sensitive measurements probe the entire film, in contrast to the XPS measurements, which are primarily sensitive to the near-surface region of the film. Analysis of the near-edge region (x-ray absorption near edge spectroscopy, XANES) were used to determine the Fe oxidation state, while the extended fine structure (EXAFS) was used to determine the Fe coordination. Spectroscopic ellipsometry measurements were performed using a J.A. Woollam variable angle spectroscopic ellipsometer over an energy range of 0.4 eV to 6.0 eV. Details of the above measurements and analysis can be found in the supplementary information. Temperature-dependent resistivity measurements were performed in a four-point probe configuration using a Quantum Design PPMS.

3. Results

Fe 2*p* core-level spectra for the Fe-doped samples are shown in Figure 1(a), along with reference spectra for LaFeO$_3$ (LFO, Fe$^{3+}$) and FeO (Fe$^{2+}$). In the case of the nominally Fe$^{3+}$-doped sample, there is a clear similarity between the spectra for the doped film and the LFO reference.[17] The spectrum for the nominally $Fe^{2+}$-doped sample shows evidence of both $Fe^{2+}$ and $Fe^{3+}$ intensity. It is possible that the $Fe^{3+}$ ions present form at the film surface due to surface oxidation during sample cooldown. However, due to low counting statistics for the Fe *2p* core level in a doped film (due to the low *x* value and the STO cap), it is not possible to perform reliable angle-resolved XPS measurements to determine if this is the case. Similar behavior has been previously observed in $Fe_2TiO_4$ spinels, with $Fe^{3+}$ present at the surface and $Fe^{2+}$ in the bulk.[18] On the other hand, given the instability of the FeO phase and its propensity to incorporate some $Fe^{3+}$ ions in the bulk,[17] it is also possible that $Fe^{3+}$ is present in the bulk of the film. For both samples, the Ti *2p* spectra (Figure 1(b)) clearly reveal $Ti^{4+}$, albeit with a slight peak broadening. Fits to the data (shown in the supplemental information) showed no evidence of a $Ti^{3+}$ $2p_{3/2}$ peak, which would lie between 456 and 457 eV for La-doped STO. However, the sensitivity to $Ti^{3+}$ is limited to ~1-2% with this technique.[19]

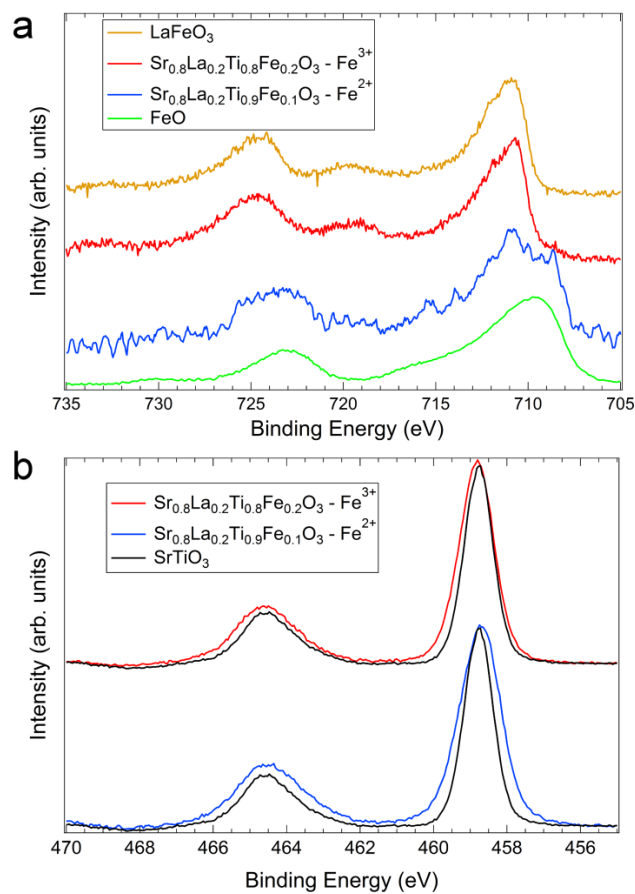

**Figure 1.** (a) Fe $2p$ x-ray photoelectron spectroscopy (XPS) data for $Fe^{2+}$ and $Fe^{3+}$ doped films with FeO and LaFeO$_3$ reference spectra; (b) Ti $2p$ XPS data for the same films with SrTiO$_3$ reference.

To further characterize the samples and evaluate the Fe XANES results for the two doped films on STO, along with LFO, FeO and Fe$_3$O$_4$ references are shown in Figure 2(a). Interpretation of the spectra is challenging, as FeO is not an ideal reference spectrum for $Fe^{2+}$ in the STO lattice. To overcome this, we examine the first derivative of the spectra from our film samples and the references, as shown in Figure 2(b). The inflection point (peak in the first derivative) along the white line rising edge is a good proxy for the Fe oxidation state.[20] We find that the inflection points are extremely well matched for the LFO reference and $Fe^{3+}$-doped sample. Meanwhile, the peak for the $Fe^{2+}$-doped sample is very close but not exactly aligned with the strongest peak of the FeO reference. The Fe$_3$O$_4$ reference exhibits two peaks in the first derivative, one near the maximum in the FeO reference and one near the LFO reference peak, consistent with the mixed valent nature of magnetite. Given the close proximity of the FeO and $Fe^{2+}$-doped peaks, it appears that the XPS data represents an overestimation of the overall $Fe^{3+}$ content. It seems likely that the $Fe^{3+}$ is concentrated near the film surface. Any estimations of the $Fe^{3+}$ content in the bulk of the film would be highly speculative, however, and therefore is not attempted.

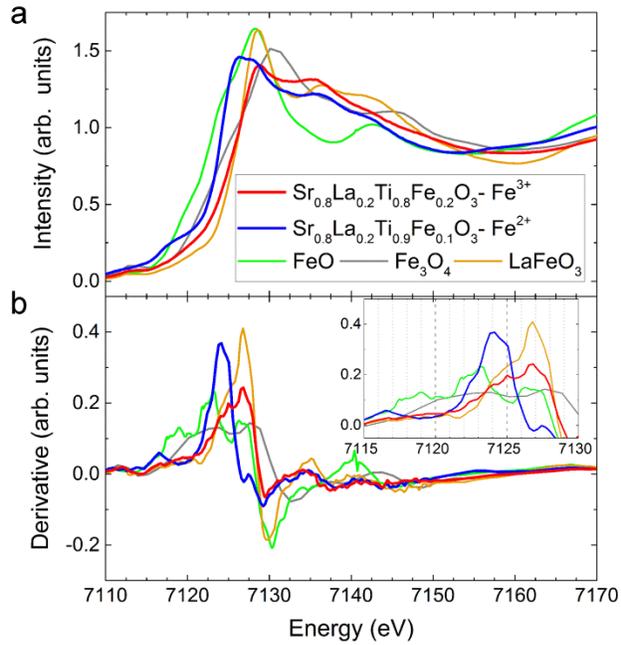

**Figure 2**. (a) Fe $K$-edge XANES spectra for $Fe^{2+}$ and $Fe^{3+}$ doped films grown on STO substrates and FeO, $Fe_3O_4$ and $LaFeO_3$ references; (b) Derivative of $K$-edge spectra with inset showing white-line region.

We also analyzed the EXAFS regions of the absorption spectra to characterize dopant lattice configurations. Radial distribution functions for the spectra, $\chi(R)$, are shown in Figure 2(b-c). Details of the fitting process can be found in the supplemental information. The EXAFS and related modeling clearly reveal that Fe substitutes for Ti in the lattice for both samples. The Fe-O bond lengths extracted from the fitting are particularly helpful in understanding the Fe spin configuration. For the $Fe^{3+}$-doped sample, the Fe-O bond length of 1.977(8) Å is between the reported value for $LaFeO_3$ (2.007(4) Å)[21] and that of $Fe^{3+}$-doped STO with no La co-dopants (1.955(5) Å)[22]. In the case of the $Fe^{2+}$-doped sample, the bond length of 2.036(14) Å is significantly less than what is expected for the high-spin $(t_{2g})^4(e_g)^2$ $Fe^{2+}$ ion as exists in the FeO phase (2.13 Å).[23] However, $Fe^{2+}$ ions in the low-spin $(t_{2g})^6(e_g)^0$ configuration would be expected to have a different Fe-O bond length than is observed in FeO, given the different $e_g$ occupancy. There is limited experimental work on low-spin $Fe^{2+}$ in oxides, but work in inorganic Fe spin-crossover molecules shows that the Fe-N bond length decreases from 2.16(2) Å in the high-spin configuration to 2.00(2) Å in the low-temperature, low-spin configuration.[24] DFT models for octahedral low-spin $Fe^{2+}$ bound to O in $H_2O$ ligands (i.e. $[Fe(H_2O)_6]^{2+}$) also found bond lengths between 1.99 and 2.03 Å.[25] These results are in good agreement with the value of 2.036(14) Å from our fit, indicating that the dopants are in the low-spin $(t_{2g})^6(e_g)^0$ state.

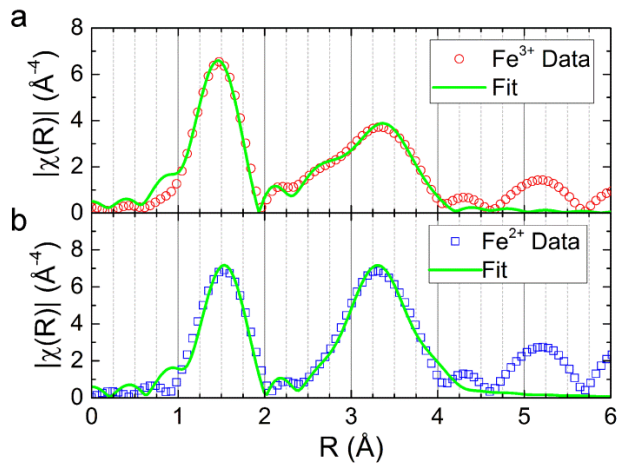

**Figure 3**. EXAFS data and fits for (a) $Fe^{3+}$-doped and (b) $Fe^{2+}$-doped samples.

To explore the electronic impact of the dopants, valence band XPS measurements were carried out immediately after film growth. Figure 4(a) shows valence band spectra for doped samples grown on STO, along with a reference STO spectrum. In the case of the $Fe^{3+}$-doped sample, the change in valence band structure is consistent with what would be expected from a solid solution of LFO and STO. A sum of LFO and STO valence band reference spectra weighted in proportion to the La and Fe doping levels (shown in the supplemental information) matches the valence band spectrum for the $Fe^{3+}$-doped film very well. The $Fe^{2+}$ sample shows a broad density of states above the O 2$p$ valence band maximum of pure STO. The intensity between ~1.5 eV and ~2.5 eV in the $Fe^{2+}$-doped spectrum can be attributed to the $Fe^{3+}$ ions near the film surface, as this region matches both our own $Fe^{3+}$-doped sample and previous observations in LTO-LFO multilayers.[12] The remaining intensity between ~1.5 eV and ~0 eV is attributable to $Fe^{2+}$ dopants deeper within the film, in agreement with previous observations and predictions.[12,13] In these works (Ref. 12 and 13), the Fe-derived density of states is extremely small at the energies where the O *2p* band lies (~3-9 eV in Figure 4(a)). This can explain the strong overlap between the STO reference spectrum and the valence band for the $Fe^{2+}$-doped sample in this region. In combination with the EXAFS results, this strongly supports the notion that the $Fe^{2+}$ dopants take on the low-spin configuration.

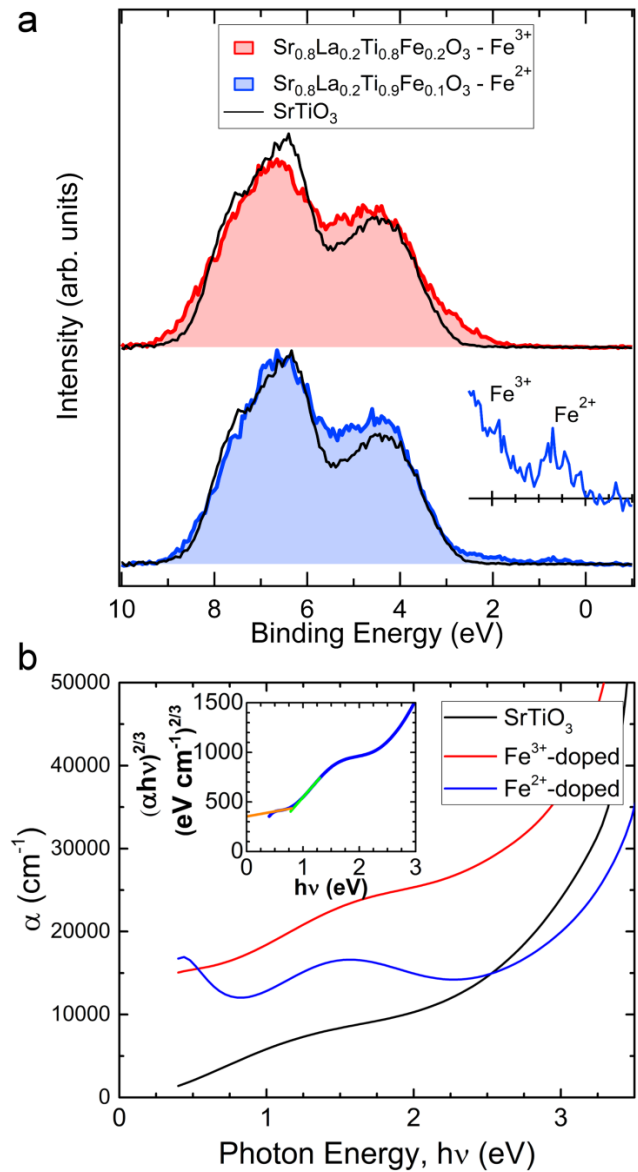

**Figure 4**. (a) Valence band spectra for $Fe^{3+}$-doped and $Fe^{2+}$-doped STO films grown on STO, and reference STO substrate; (b) Optical absorption data for doped films and STO reference grown on $DyScO_3$; (c) Schematic band diagram for both dopant configurations.

Given the presence of Fe $t_{2g}$ electronic density near the Fermi level, the optical absorption spectrum of the $Fe^{2+}$-doped film can provide important insights into the electronic structure of the material and its potential for solar applications. Spectroscopic ellipsometry measurements were performed on the films grown on DSO and were fitted to obtain the absorption coefficient, $\alpha$, as shown in Figure 4(b). Details of this analysis are provided in the supplemental information. The $Fe^{2+}$-doped film shows a clear absorption peak at energies between ~0.7 eV and ~2.0 eV. The STO and $Fe^{3+}$-doped samples show mutually similar features with greater overall absorption in the $Fe^{3+}$-doped spectrum, most likely due to defect-induced states. These features are distinct from what is observed in the $Fe^{2+}$-doped sample. This absorption band represents a low-energy optical transition rarely, if ever, observed in STO and doped STO films. Such an absorption band could have significant applications for photocatalysis and photovoltaics if it can be controlled through the growth process.

The $Fe^{2+}$-doped film also shows a Drude response at low energies consistent with high free carrier densities, as has been observed in La-doped STO.[26] The free carriers may be due to one of two factors: the mixture of $Fe^{2+}$ and $Fe^{3+}$ which produces room-temperature hopping conductivity analogous to Ti-doped $Fe_2O_3$,[27] or thermal carrier activation into the Ti $3d$ conduction band due to a band gap that is not significantly greater than $kT$ at room temperature. The Tauc formula for a dipole-forbidden direct gap Fe $3d \rightarrow$ Ti $3d$ transition was used to estimate the band gap by linearly extrapolating $(\alpha h\nu)^{2/3}$ vs. $h\nu$ through the intersection with the background intensity from the Drude response.[29,30] Using this method (see inset of Figure 4(b)), we determine a direct forbidden band gap of 0.83(5) eV. The presence of free carriers makes precise determination of the optical band gap problematic due to the Burstein-Moss band filling phenomenon, which overestimates the true gap in degenerately doped systems,[28] and thus this value is expected to be an overestimate of the true gap. As has been noted, Tauc plot projections may be somewhat subjective and prone to systematic error,[30] so the systematic uncertainty in our estimate is somewhat larger than the uncertainty in the linear extrapolation.

Temperature-dependent resistivity measurements were performed on both doped samples grown on DSO to examine the behavior of the free carriers observed in the $Fe^{2+}$-doped sample. The graph of resistivity, $\rho$, as a function of temperature, $T$, for the two samples is shown in Figure 5(a). The $Fe^{3+}$ sample shows strong insulating behavior with temperature. Meanwhile, the $Fe^{2+}$ sample undergoes a metal-semiconductor transition at ~250 K as determined by the zero crossing of the derivative $d\rho/dT$. At temperatures below 125 K, the semiconducting behavior can be modeled very well by the Efros-Shklovskii (E-S) variable range hopping (VRH) model[31]:

$$\rho(T) = \rho_0 e^{(\frac{T_0}{T})^{1/2}} \quad (1),$$

where $T_0$ and $\rho_0$ are constants. A linear fit to the data $\log(\rho)$ vs. $1/T^{1/2}$ is shown in Figure 5(b). Models of both 2-dimensional ($\log(\rho)\sim(T_0/T)^{1/3}$) and 3-dimensional ($\log(\rho)\sim(T_0/T)^{1/4}$) Mott VRH[32] in the low temperature regime were non-linear, showing poor agreement with the data. These models are shown in the supplemental information. Near room temperature, the sample becomes metallic, with $\rho(T)$ following the classical three-dimensional Fermi liquid $T^2$ dependence,[33] as shown by a fit to $\rho$ vs. $T^2$ in Figure 5(c).

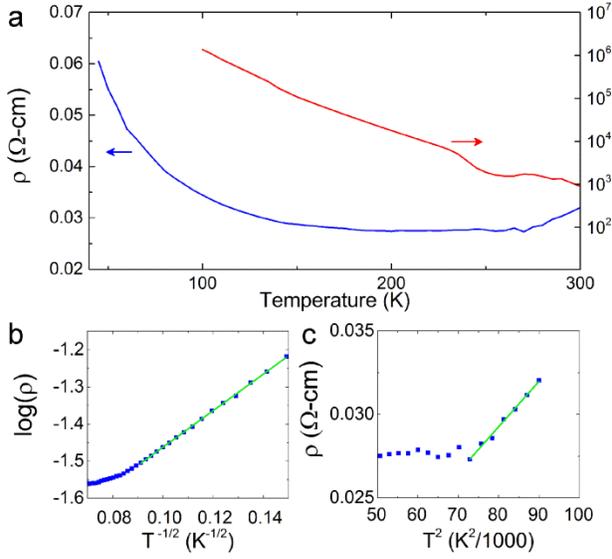

**Figure 5**. Temperature-dependent resistivity measurements of $Fe^{2+}$-doped (blue) and $Fe^{3+}$-doped (red) samples grown on $DyScO_3$ (a) with linear fits to (b) Efros-Shklovskii variable range hopping at low temperatures and (c) metallic $T^2$ resistivity at high temperatures.

## 4. Discussion

The conversion from high-spin (HS) to low-spin (LS) $Fe^{2+}$ is attributable to an increase in the crystal field splitting energy, 10Dq, as was predicted in the case of LFO-LTO interfaces.[12] This is shown schematically in Figure 6(a). As the value of 10Dq increases, there is sufficient energy to overcome Hund's exchange and produce a LS $Fe^{2+}$ ion. This is largely a function of the coordination of the $Fe^{2+}$ ion within the STO lattice, where the octahedral volume is far less than in the case of rocksalt FeO with HS $Fe^{2+}$. Thus, the measured Fe-O bond length is a key indicator of LS $Fe^{2+}$.

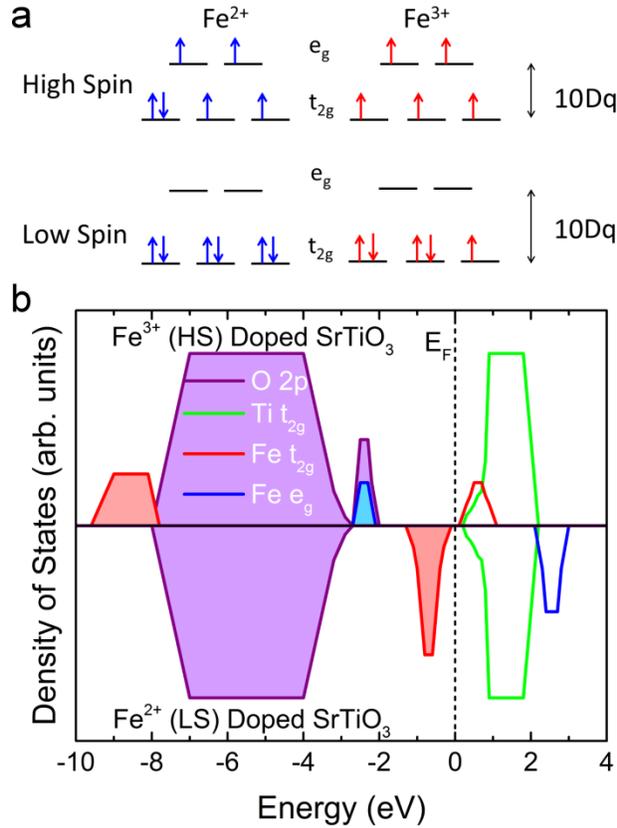

**Figure 6**. Schematic diagrams of (a) high and low spin electronic configurations of Fe oxidation states and (b) valence band diagram of Fe-doped $SrTiO_3$ films studied in this work.

The results of the spectroscopic ellipsometry and resistivity measurements for the $Fe^{2+}$-doped sample are rather unusual, as they indicate that the film is metallic at room temperature and semiconducting at low temperature. The low temperature E-S VRH conductivity is not surprising in such a system, as the model is based on the assumption of a non-uniform density of states very close to the Fermi level,[31] which we have observed in our valence band XPS measurements. E-S VRH has been observed in other oxides, including $In_xO_y$[34], $NbO_{2+\delta}$ films[35] and $VO_x$ films[36]. In the case of narrow gap oxides, Goodenough predicted that mixed oxidation states would act as hopping centers,[37] as is the case for each of the mixed valence oxides where E-S VRH is typically observed. Given that our nominally $Fe^{2+}$ doped films are known to have some $Fe^{3+}$ ions present at least near the film surface, this is the most likely mechanism for the hopping conductivity that is observed.

The measured optical gap for the Fe $3d \rightarrow$ Ti $3d$ transition of ~0.8 eV is somewhat greater than the ~0.5 eV value that was predicted via DFT previously.[12] As noted above, the Burstein-Moss effect due to free carriers may explain some or all of this discrepancy. Burstein-Moss shifts on the order of a few tenths of an eV have been observed in La-doped STO via ellipsometry previously.[26] Thus, while we cannot provide a precise estimate of the optical gap, our results do not necessarily contradict previous predictions for the system. Given that the gap is significantly greater than $kT$ at room temperature, thermal excitation from $Fe^{2+}$ into unoccupied $Fe^{3+}$ defect states is the most likely mechanism for the conductivity that we observe. At sufficiently high temperatures the number of filled defect states is likely sufficient to form a delocalized band, producing metallic behavior.

While degenerately-doped materials are interesting in their own right, photovoltaic and photocatalytic applications demand somewhat wider band gaps. The ideal optical band gap for photovoltaic and photocatalytic materials is reported to be ~1.5 eV to produce strong visible light absorption.[30,38] Thus, to fully realize applications of $Fe^{2+}$ dopants using this scheme, a technique to raise the bottom of the conduction band in STO is needed. Recent work has indicated that this may be achieved in solid solutions of STO and $SrZrO_3$.[39] $SrZrO_3$ is isovalent with STO, but has a larger band gap of 5.6 eV.[40] Substitution of Zr ions for Ti has been shown to produce a wide, tunable range of band gaps by removing low lying Ti $3d$ states.[39] Using this approach, it may be possible to raise the bottom of the conduction band to reduce free carrier concentrations and control the absorption energies induced by $Fe^{2+}$ dopants.

5. **Conclusion**

In summary, we have demonstrated that by co-doping $SrTiO_3$ with La on the A site and Fe on the B site, we can control the oxidation state of the dopant ions and engineer the valence band structure. As expected, by doping one $La^{3+}$ ion for every Fe ion, we stabilize substitutional $Fe^{3+}$ and achieve a valence band consistent with that of an LFO-STO alloy. We are also able to stabilize the low-spin $Fe^{2+}$ state by doping two $La^{3+}$ ions for every Fe ion in the system. Our results agree with theoretical predictions for both the $Fe^{2+}$-O bond length and the predicted valence band structure. An optical absorption band in the near infrared regime originates from the Fe $t_{2g}$ states that lie near the Fermi level of the material. Measurements of the energy for this optical transition suggest that the gap for this transition is ~0.8 eV, but are complicated by the presence of free carriers which artificially increase the value of the gap due to the Burstein-Moss effect. Previous theoretical predictions of a band gap of 0.5 eV are reasonable given the systematic uncertainties. Further work to lift the bottom of the conduction band via Zr substitution for Ti may help to realize photovoltaic and photocatalytic applications for these materials by increasing the band gap towards the ideal value of ~1.5 eV.

**Acknowledgment**

R.B.C. was supported by the Linus Pauling Distinguished Postdoctoral Fellowship at Pacific Northwest National Laboratory (PNNL LDRD PN13100/2581). S.A.C. and M.E.B. were supported at PNNL by the U.S. Department of Energy, Office of Science, Chemical Sciences, Geosciencs, and Biosciences, under Award No. 48526. The PNNL work was performed in the Environmental Molecular Sciences Laboratory (EMSL), which is a national science user facility sponsored by the Department of Energy's Office of Biological and Environmental Research and located at Pacific Northwest National Laboratory. PNC/XSD facilities at the Advanced Photon Source, and research at these facilities, are supported by the U.S. Department of Energy−Basic Energy Sciences, the Canadian Light Source and its funding partners, the University of Washington, and the Advanced Photon Source. Use of the Advanced Photon Source, an Office of Science User Facility operated for the U.S. Department of Energy (DOE) Office of Science by Argonne National Laboratory, was supported by the U.S. DOE, under Contract No. DE-AC02-06CH11357. The authors gratefully acknowledge helpful discussions with Dr. Tim Droubay of PNNL.

# Infrared Optical Absorption in Low-spin $Fe^{2+}$-doped $SrTiO_3$

Ryan B. Comes,[1] Tiffany C. Kaspar,[1] Steve M. Heald,[2] Mark E. Bowden,[3] and Scott A. Chambers[1]

[1]Physical and Computational Sciences Directorate, Pacific Northwest National Laboratory, Richland, WA 99352, United States

[2]Advanced Photon Source, Argonne National Laboratory, Argonne, IL 60439, United States

[3]Environmental Molecular Sciences Laboratory, Pacific Northwest National Laboratory, Richland, WA 99352, United States

Supplemental Online Information

*Film Growth*

La, Fe co-doped $SrTiO_3$ (SLTFO) films were grown on $SrTiO_3$ (STO) (001) and $DyScO_3$ (DSO) (110) substrates using oxygen-assisted molecular beam epitaxy from four effusion cells. DSO is orthorhombic, so the (110) orientation is equivalent to the pseudocubic (001) axis. Prior to growth, the fluxes of the cells were calibrated to within 5% precision using a quartz crystal microbalance. The Sr:Ti ratio was then further refined to within 1-2% through a shuttered growth scheme by analysis of reflection high energy electron diffraction (RHEED) oscillations.[1] This approach has been successfully employed for La, Cr co-doped films previously and is discussed in detail in that work.[2] The doped films were then grown at a substrate temperature of 700° C at a pressure of $3\times10^{-6}$ Torr. For the $Fe^{3+}$ doped films, activated atomic oxygen was supplied using an ECR plasma source, while the $Fe^{2+}$ films were grown in molecular oxygen to prevent overoxidation to $Fe^{3+}$. The La:Fe ratio was tuned to be slightly La deficient (~5%) to prevent excess La ions from unintentionally introducing carriers in the samples. The films were capped with a 2 unit cell STO layer to further protect the film during cooldown. RHEED patterns after growth for the films grown on DSO are shown in Figure S1.

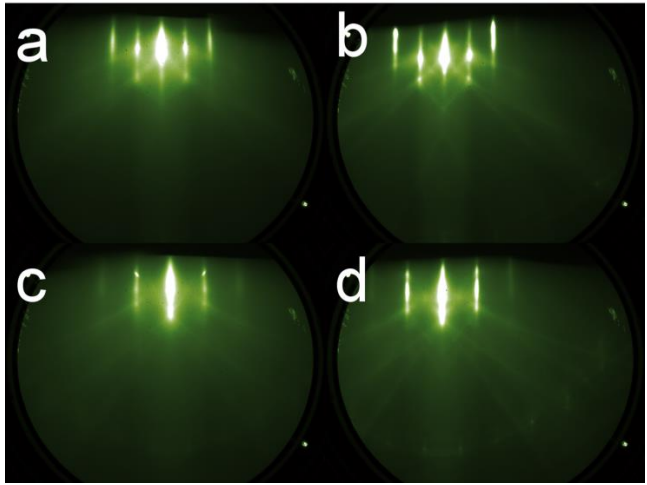

**Figure S1:** RHEED images along (10) for (a) $Fe^{3+}$-doped and (b) $Fe^{2+}$-doped films and (c-d) along (11) for the same films.

*X-ray Diffraction*

Out-of-plane x-ray diffraction and reciprocal space maps about the pseudocubic (103) peaks were performed on the samples to determine the lattice parameters and estimate the structural quality of the materials. Out-of-plane scans for the SLTFO films and STO reference grown on $DyScO_3$ are shown in Figure S2. The measured out-of-plane lattice parameters are shown in Table S1.

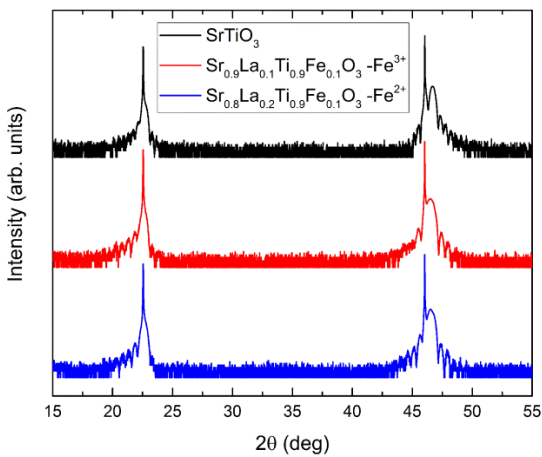

**Figure S2:** Out-of-plane x-ray diffraction data for films grown on DyScO3.

Table S 1: Fitted out-of-plane lattice parameters for samples grown on DyScO$_3$.

| Sample | Out-of-plane lattice parameter (Å) |
| --- | --- |
| SrTiO$_3$ | 3.888(8) |
| Sr$_{0.9}$La$_{0.1}$Ti$_{0.9}$Fe$_{0.1}$O$_3$ | 3.902(16) |
| Sr$_{0.8}$La$_{0.2}$Ti$_{0.9}$Fe$_{0.1}$O$_3$ | 3.898(16) |

*X-ray Photoelectron Spectroscopy*

Fits were performed to the Ti $2p$ spectrum for the Fe$^{2+}$-doped sample grown on STO to verify that there was no measurable intensity of Ti$^{3+}$ in the material. These results are shown below in Figure S3. After background subtraction with a Shirley background, the $2p^{3/2}$ and $2p^{1/2}$ peaks were fitted with Voigt functions and the satellites were fitted with Gaussians. A slight asymmetry in the $2p^{3/2}$ peak is observed, but there is no error in the fit in the 456-457 eV range where Ti$^{3+}$ intensity is generally observed.

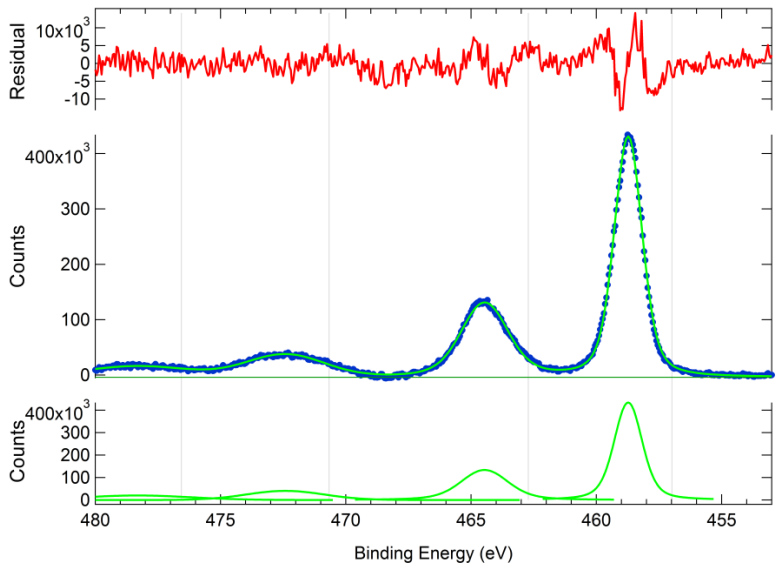

**Figure S3: Raw data (blue), fitted spectrum (green) and residual (red) for the Ti $2p$ peak in the Fe$^{2+}$-doped film grown on STO.**

To model the valence band spectrum of the $Fe^{3+}$-doped sample, a summation spectrum was composed of STO and $LaFeO_3$ (LFO) valence band references. These two were then summed together in propotion to the amount of La and Fe doping in the material after alignment to produce a composite valence band shown in Figure S4. The composite spectrum matches the observed features very well, suggesting that the $Fe^{3+}$ dopants in this material take on a similar electronic spin configuration to that of pure LFO.

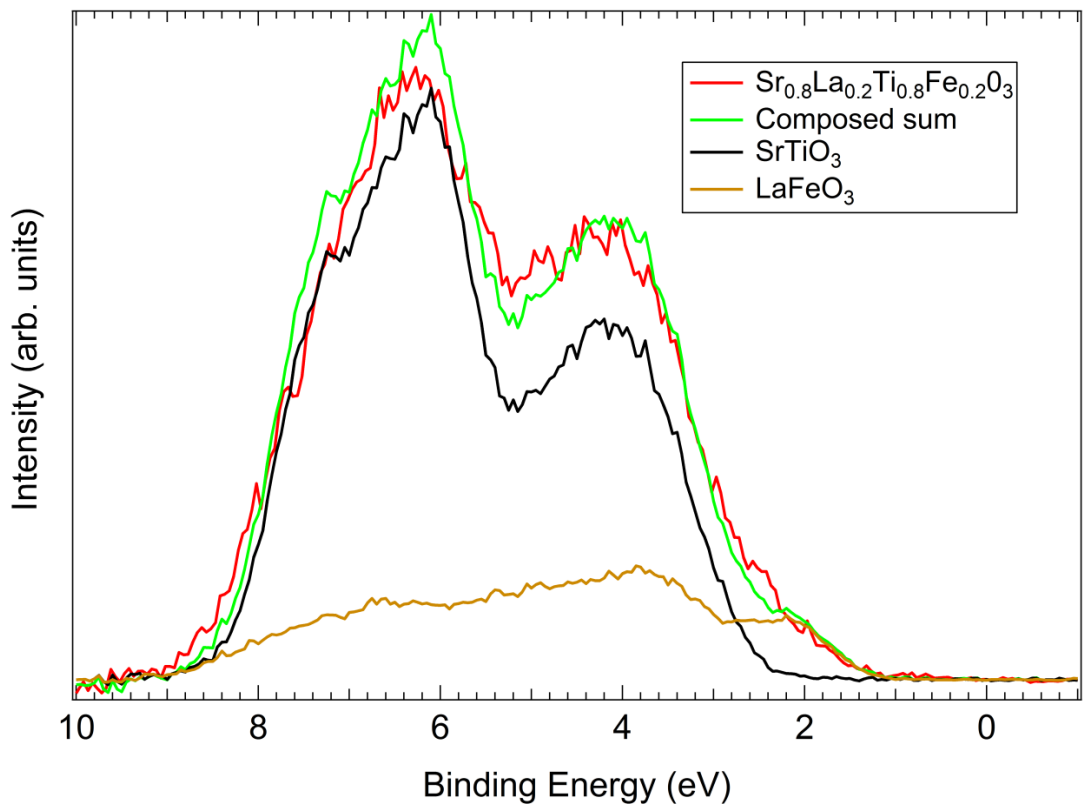

**Figure S4: Composite valence band model of the $Fe^{3+}$-doped valence band. $SrTiO_3$ and $LaFeO_3$ bands are shown in proportion to their contribution and the measured valence band is normalized to align the low energy falling edge.**

*X-ray Absorption Spectroscopy Measurements*

Measurements were performed on the Fe *K*-edge at the Advanced Photon Source on beamlines 20-BM using a Si (111) monochromator with energy resolution of about 0.8 eV. Energy calibration was done by setting a Fe metal standard edge to 7110.75 eV. Scans were taken with the x-ray polarization perpendicular to the surface of the films. The samples were set at a small angle (5-7°) to the focused beam and spun around the sample normal to avoid interference from sample Bragg reflections. The data was analyzed using the Demeter XAS software suite.[3] The fitting range was for radial distances of 1-4 Å and in *k*-space for 2-9.5 Å$^{-1}$. Single and multiple scattering paths from the FEFF code with effective radii of 4.4 Å and less were used in the models. Fits were performed to determine the energy shift, $E_0$, overall amplitude, $S_0^2$, nearest neighbor O, Ti, and *A* site distances, and Debye-Waller factors for the nearest neighbors. The number of nearest neighbor La ions, $N_{La}$, was varied manually to account for bonding between La and Fe ions in the lattice during the growth process to minimize the R factor of the fit.[2] The results of the fits for both the $Fe^{2+}$- and $Fe^{3+}$-doped films are shown below in Table S2.

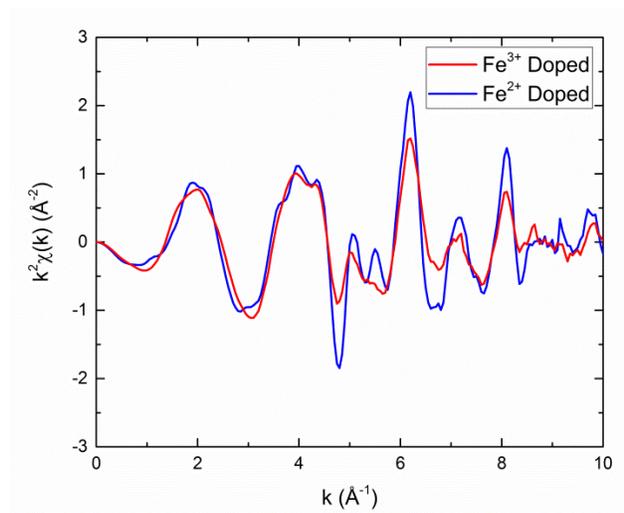

**Figure S 5: Raw reciprocal space EXAFS data for the two samples modeled.**

**Table S2: EXAFS fitting results.**

| Sample (STO substrate) | $\Delta E_0$ (eV) | $S_0^2$ | Fe-O Bond Length (Å) | Fe-O Debye-Waller Factor (Å$^2$) | Fe-Ti Bond Length (Å) | $N_{La}$ | R factor |
|---|---|---|---|---|---|---|---|
| **$Sr_{0.8}La_{0.2}Ti_{0.8}Fe_{0.2}O_3$** | -5.7(9) | 0.69(8) | 1.977(8) | 0.0075(17) | 3.93(2) | 1.6 | 0.010 |
| **$Sr_{0.8}La_{0.2}Ti_{0.9}Fe_{0.1}O_3$** | -2.1(1.3) | 0.71(13) | 2.036(14) | 0.0065(27) | 3.95(2) | 3.0 | 0.020 |

*Spectroscopic Ellipsometry Measurements*

Variable angle spectroscopic ellipsometry measurements were made using a J.W. Woollam VASE system over an energy range of 0.4-6.0 eV and at angles of 50, 60 and 70 degrees. The data were fit to obtain the energy dependent index of refraction, *n*, and extinction coefficient, *k*. The absorption coefficient, *α*, was derived using the formula:

$$\alpha = \frac{4\pi k}{\lambda} \quad (S1),$$

where $\lambda$ is the photon wavelength.

*Electronic Transport*

Models of the low-temperature transport assuming Mott variable range hopping (VRH) in two and three dimensions are shown below in Figure S6.[4] The temperature dependence of the conductivity should be linear on the scales presented for good agreement with the models, which is clearly not the case.

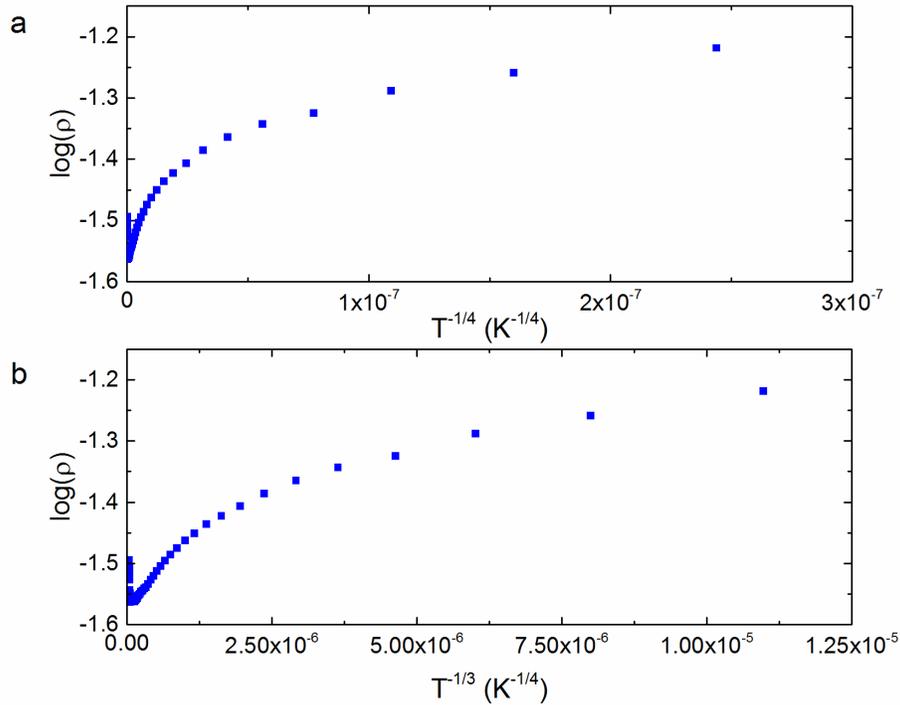

**Figure S6:** a) Logarithmic resistivity plotted against $T^{-1/4}$ for 3D VRH model; b) Logarithmic resistivity plotted against $T^{-1/3}$ for 2D VRH model. Data should be linear on these scales for accurate model.